\documentclass[a4paper,12pt,oneside]{article}
\usepackage{amsmath}
\usepackage{wrapfig}
\usepackage{graphicx}
\usepackage{lscape}
\usepackage[cp1250]{inputenc}
      
\title{Relations between organisms and the environment in the ageing process}
\author{Przemys{\l}aw Biecek, Katarzyna Bońkowska, Stanis{\l}aw Cebrat
\\
\small\text{Department of Genomics}
\\
\small\text{Wroc{\l}aw University, ul. Przybyszewskiego 63/77, 51-148 Wroc{\l}aw, Poland, }
\\
\small\text{cebrat@microb.uni.wroc.pl }\\
}

\date{}

\begin{document} 
\maketitle

\begin{abstract}
We have modified the sexual Penna model by introducing the fluctuating environment and fluctuations representing physiological functions of individuals.  Additionally, we have introduced the mother care corresponding to the protection against the deleterious influence of the environment, the learning capacity of individuals corresponding to their immunity and adaptation to the environment fluctuations and the other risk factors appearing at puberty. Each of the above mentioned elements influences mainly the survival of newborns and young individuals while genetic defects accumulated in the genomes increase the noise of individuals and were responsible for higher mortality of older individuals. All these modifications enable precise fitting the age structure of the simulated populations to the age distributions of human populations.
\end{abstract}

\section{Introduction}
In 1952 Medawar formulated the theory of ageing called the “mutations accumulation theory” \cite{Medawar}. He argued that: if there is a random death killing the individuals in a population independently of their age, the sizes of groups of individuals at a given age decrease exponentially with age; if all individuals keep the same reproduction potential per time unit during the whole lifespan, the reproduction potential of age groups decreases also with age. In consequence, selection pressure on the oldest individuals will decrease which allows the accumulation of defects in the genes indispensable for surviving the later periods of the lifespan. Therefore, the accumulation of defective genes which are expressed after the minimum reproduction age should be observed and these defects would be responsible for the genetic program of ageing.

In 1995 Penna described a simple computer model, in which he assumed that genes in genomes are switched on chronologically \cite{Penna}. He observed that in the fraction of genes switched on before the reproduction age much less mutations were accumulated than in the fraction of genes switched on after the minimum reproduction age. It seems to be obvious that defects expressed before the reproduction kill the individual and they cannot be transferred to the offspring. If defects are expressed after the minimum reproduction age, they can be inherited by the offspring and the chance of being transferred is still higher for genes expressed later during the lifespan. For description of the original model, its different versions and results of simulations of many demographic phenomena see \cite{Stauffer}, \cite{JIM}.

In the standard Penna model, the size of a population was controlled by Verhulst factor: $V=1-N_t/N_{\max}$, where $V$ describes the probability of survival for the individual independently of its age,
$N_t$ corresponds to the actual size of the population and $N_{\max}$ is called
the maximum capacity of the environment. It was the Verhulst factor which introduced the random death, assumed by Medawar, into the model. But in the other version of the model, the Verhulst factor determined the survival probability of the newborns only \cite{Martins} in fact it controlled the birthrate. Like in nature, the reproduction probability and the survival of newborns depend on the resources availability and overcrowding. In simulations where the Verhulst test was set for newborns only, the random death of older individuals disappeared. In such models the accumulation of mutations in the fraction of genes expressed later during the lifespan is still observed. Furthermore, the genetic pool structure of such populations is better (lower fraction of defective genes). The results indicate that the random death during the lifespan in the population is not necessary for generating the nonrandom distribution of defective genes in the genomes, neither for emerging the genetically determined ageing. Using this model it is possible to recover in the computer simulations the age structure of populations which fits the real human populations \cite{Niewczas}. Furthermore, it is possible to simulate the changes in the age structure of human populations and to predict demographic changes for the near future \cite{Aga}, \cite{Kasia}. Nevertheless, such a modeling renders good results for fractions of populations in the reproduction ages, only. The age structure of the part of population before the reproduction age does not correspond to the real demographic data.

The standard Penna model produces the low but constant fraction of defective genes in the set of genes expressed before the reproduction age. An important parameter of the model is the threshold number of defective phenotypes $T$. If this number is reached by an individual it has to die because of its genetic status. An individual cannot die before it reaches the threshold $T$ e.g. for $T=3$ the genetic death is impossible during the first two time units (Monte Carlo steps). Many authors tried to solve the problem of “the babies death” in the Penna model. They introduced lower threshold $T$ for babies \cite{Berntsen}, higher number of genes switched on in one time unit or introduced genes switched on in the periods between conception and birth \cite{Kurdziel}. All these modifications were biologically legitimated but there was no place in the model for more or less random relationships between individuals and environment. Such relations were introduced in the recent versions of the model \cite{Biecek}, \cite{learning} where both, state of an individual and state of the environment fluctuate. If the sum of both fluctuations passes the limit set for homeodynamics of the organism then it has to die. If the babies are more sensitive for such a noise, their mortality is higher.

This phenomenon wass very well seen in the ancient European human populations when many newborns and young children died because of infections. Recently, due to the intensive medical care, the mortality of newborns dropped. In the simulations described in this paper we have introduced some parameters mimicking the mechanisms which rescue the babies from the life threatening conditions.

\section{The standard Penna model}

In the Penna diploid, sexual model each of $N$ individuals of the population is represented by a genome composed of two bitstrings (haplotypes). Bits set for $0$ represent correct genes, bits set for $1$ represent defective genes. Two bits at the same position in the bitstrings (locus) represent alleles. If both alleles are set for $1$ then the phenotype determined by the locus is defective (i.e. defective genes are recessive). Loci are switched on chronologically. Organism dies if the declared number of defective loci (threshold $T$) has been switched on. The organism can reproduce if it survives until the age $R$ (the reproduction age). During reproduction, each bitstring is copied and mutations are introduced into randomly chosen loci with a declared probability ($0$ is replaced by $1$, bit set for $1$ stays $1$). Two new bitstrings recombine with a declared probability by exchanging their arms in the randomly chosen point. Each bitstring after these processes is a gamete. The offspring is produced by joining a gamete generated by a female with a gamete produced by a male. In the sexual model each newborn is declared to be female or male with the equal probability. The example of mortality curve and the distribution of defective genes generated in the standard Penna model simulations are shown in Fig. 1.

\section{The noisy Penna model}

For further extensions of the model we use only the diploid sexual Penna model described in the above section. In this standard model Verhulst factor controls the birthrate and there are no random deaths of organisms later during the lifespan. In the standard Penna model individuals die because of genetic death when they reach the threshold $T$ of the expressed defective phenotypes, only. In the noisy version of the model there is no declared threshold $T$. Instead, we have introduced the fluctuations of the state of organisms. The energy of fluctuations increases with the number of switched on defective loci \cite{Biecek}. 
Thus the model is
\begin{equation}
I_i(t) = E(t) + P_i(t),
\end{equation}
where 
$E(t) \sim \mathcal N(\mu_{E(t)},\sigma^2_e)$ corresponds to the fluctuations of environment 
in time $t$ while $P_i(t) \sim \mathcal N(\mu_{P_i(t)},\sigma^2_i(t))$ corresponds to the inner fluctuations of individual $i$ in time $t$.
In the simplest case the expected value of both fluctuations is $\mu_{P_i(t)} = \mu_{E(t)} = 0$ and the energy of fluctuations of the state of individual depends on its number of defective loci $g_i(t)$ expressed till time $t$
\begin{equation}
\sigma^2_i(t) = \sigma^2_0 + g_i(t)\sigma_d^2.
\end{equation}

Both models produce very similar results of simulations with characteristic gradient of defective genes expressed after the minimum reproduction age and very low mortality of the youngest individuals. The only difference between the two models concerns the mortality of the individuals during the first two time units. For threshold $T=3$ in the standard model there are no genetic deaths during the first two time units (see Fig. 1). In the noisy model organisms may die even before the expression of any defect because of fluctuations (see Fig. 2).

\begin{figure}
	\centering
		\includegraphics[width=0.8\textwidth]{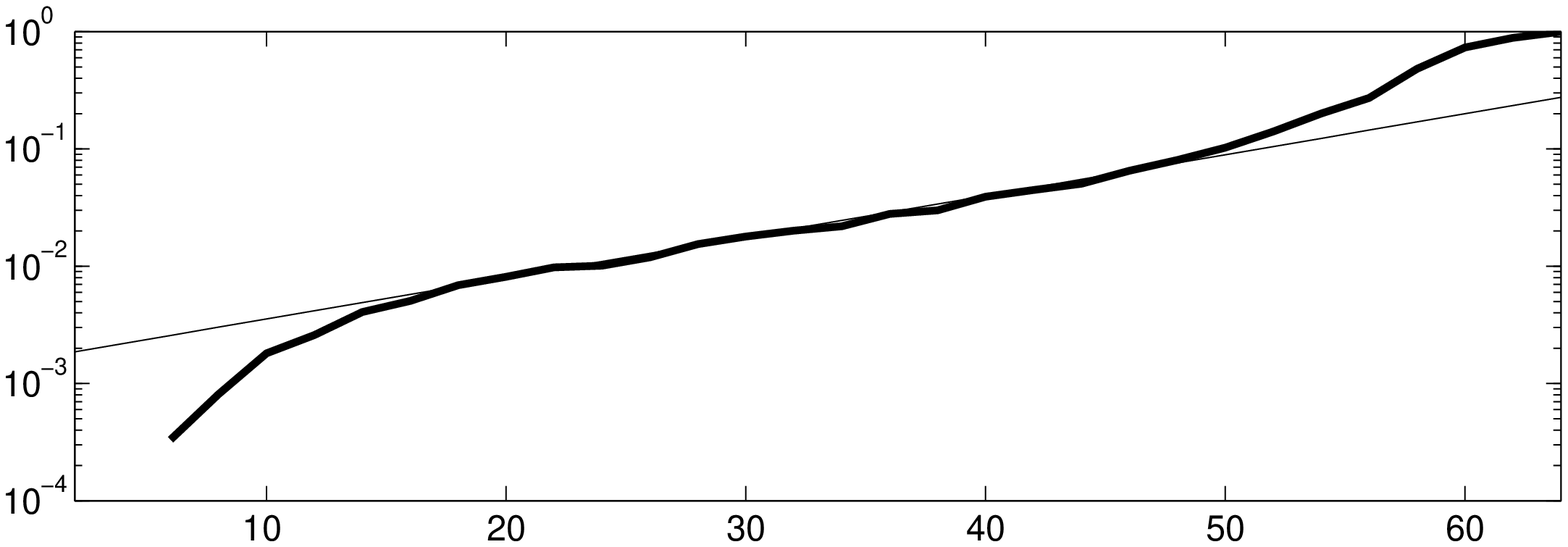}
		\includegraphics[width=0.8\textwidth]{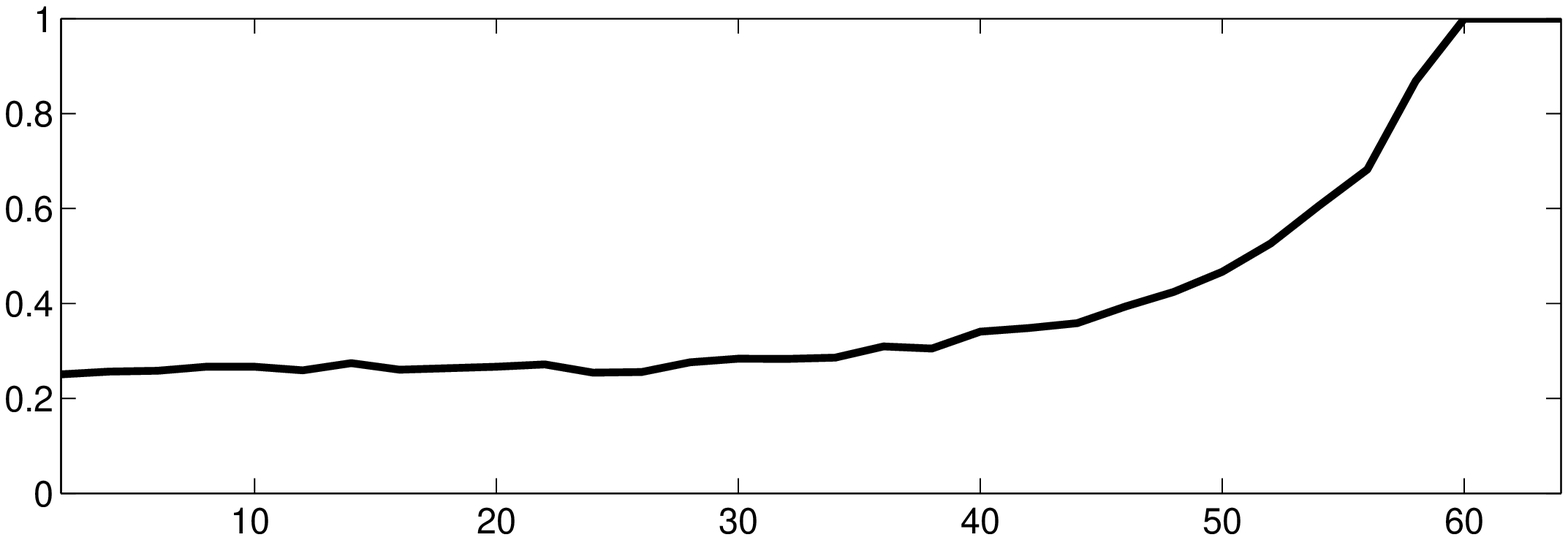}
		\caption{ The mortality curve (upper plot) and frequency of defective genes (lower plot) in the standard Penna model. Note the logarithmic y-axis of mortality plot, x-axis is scaled in the arbitrary age units.}
	\label{fig:penna}
\end{figure}

\begin{figure}
	\centering
		\includegraphics[width=0.8\textwidth]{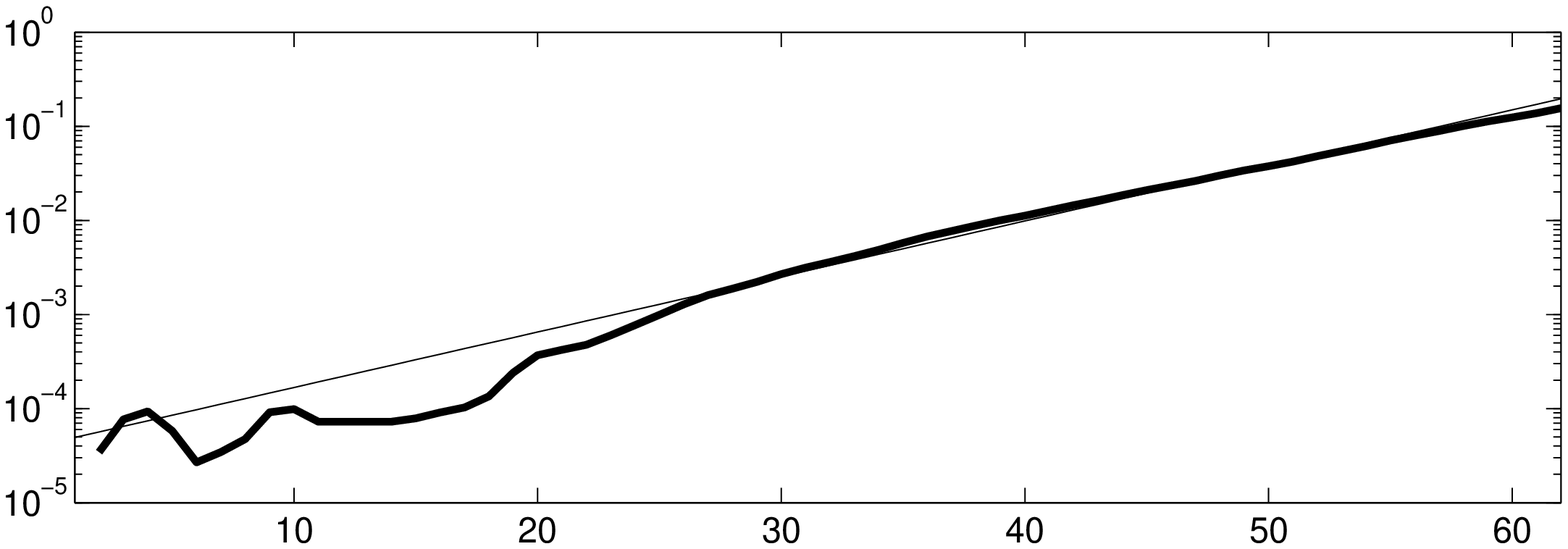}
		\includegraphics[width=0.8\textwidth]{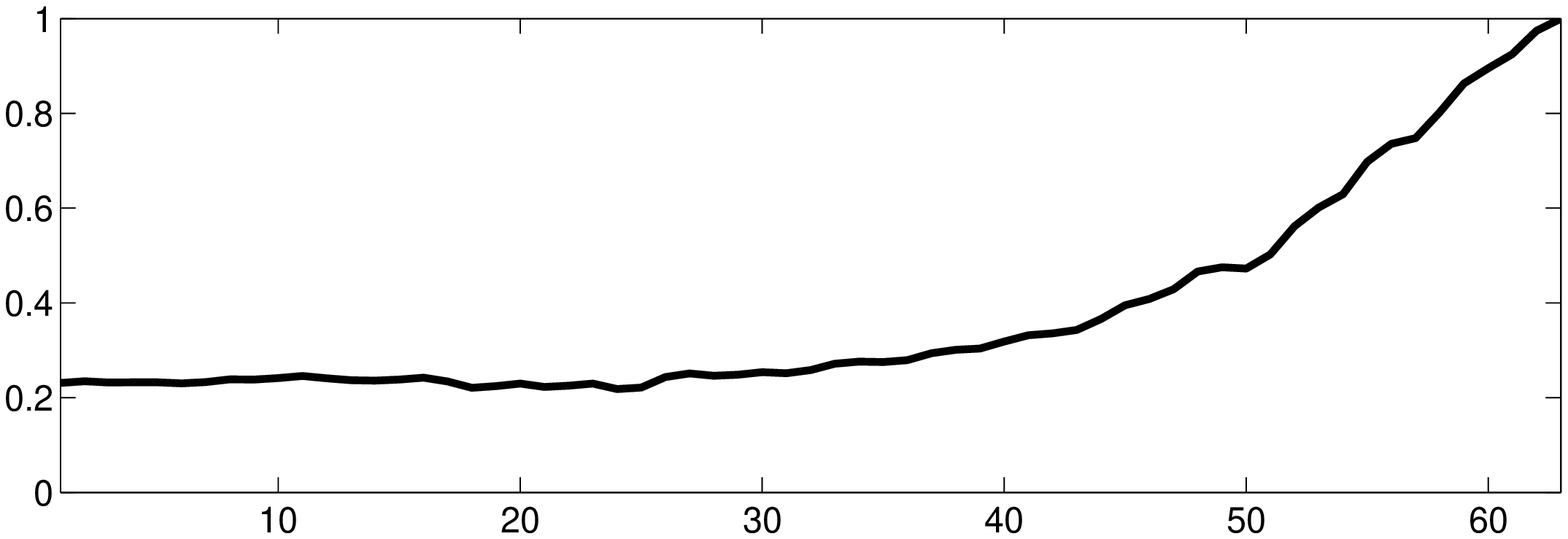}
		\caption{The mortality curve (upper plot) and frequency of defective genes (lower plot) in the noisy Penna model.}
	\label{fig:defekty}
\end{figure}

\section{Mother care}

In the noisy Penna model the state of environment affects all individuals regardless of their age.
We model the mother care as some kind of protection of the babies against the influence of fluctuations of the environment during the first periods of their lives. 
So, newborns are affected by states of the environment decreased by 
$$
\rho(i,t) = 1 - exp(-(age(i,t)+1)/\lambda_{MC}).
$$
The individual dies if $\rho(i,t) E(t) + P_i(t) > F$.
After $\lambda_{MC}$ steps the effect of mother care is negligible while in the early stages of life it is significant. 

We call this ,,mother care'' to stress that this effect influences the very first periods of life but it could be the proper feeding of newborns with mother’s milk as well as intensive medical care.
In Fig. 3 we present results for $\lambda_{MC}=4$. The fraction of defects is similar for both models.
 
\begin{figure}
	\centering
		\includegraphics[width=0.8\textwidth]{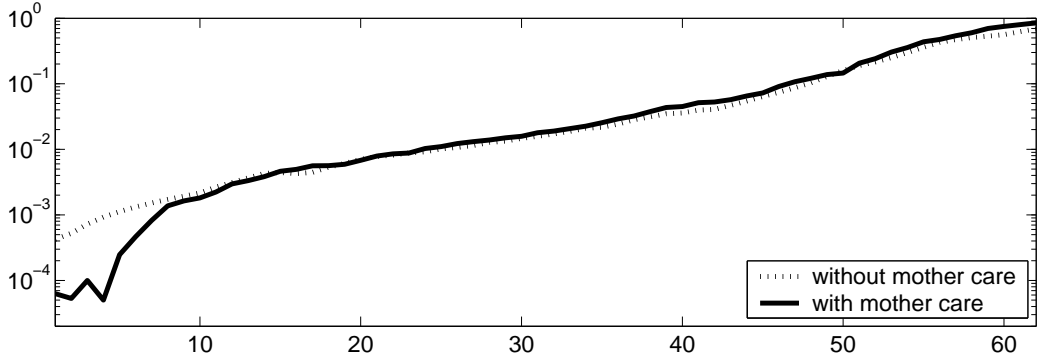}
		\caption{Mortality curves for the population with and without mother care.}
	\label{fig:addRisk}
\end{figure}

\section{Adaptation to the environmental conditions}

In the noisy Penna model the energy of fluctuations of individual is a sum of its inner noise and the noise of environment. The impact of these two components on the state and evolution of population is different. The personal component is non-correlated and independent for each individual while the environmental noise is the same for each individual. That is the reason why the reactions of individuals for the environmental fluctuations are diversified. 

In the version of the noisy model presented above, the fluctuations have Gaussian distribution with average $\mu_{E(t)}0$. Now, we introduce a signal into the expected state of environment. The signal $\mu_{E(t)}$ is a periodical function with period D, thus $\mu_{E(t)} = \mu_{E(t+D)}$. Individuals know that the signal is periodical, and are equipped with a mechanism of learning the signal. They estimate each component of the signal by weighted average of state of environment in survived periods. In the more formal way the learning mechanism affect the expected state of individual fluctuations
$$
\mu_{P_i}(t) = \sum_{j=1}^{\infty} L(i,t-j*D) w_j E(t- j*D)
$$
where $L(i,t) = 1$ if individual $i$ lived at time $t$ and $0$ otherwise while weights $w_j$ are 
$$
w_j = e^{-(j-1)/\lambda} - e^{-j/\lambda} \quad.
$$

This adaptation mechanism allows reducing the mortality in case when individual have learned the periodical signal. Results for different $\lambda$ are presented on Fig. 4. 
It is also observed in real populations, that mortality of newborns is higher than of a bit older individuals. 
The results depend on the maximal signal value $\mu_{E(t)}$ and do not depend on the form of the periodic function, thus results for constant $\mu_{E(t)}=A$ are similar to those obtained with $\mu_{E(t)}=A \sin(t/\pi)$ (results not shown).

\begin{figure}
	\centering
		\includegraphics[width=0.8\textwidth]{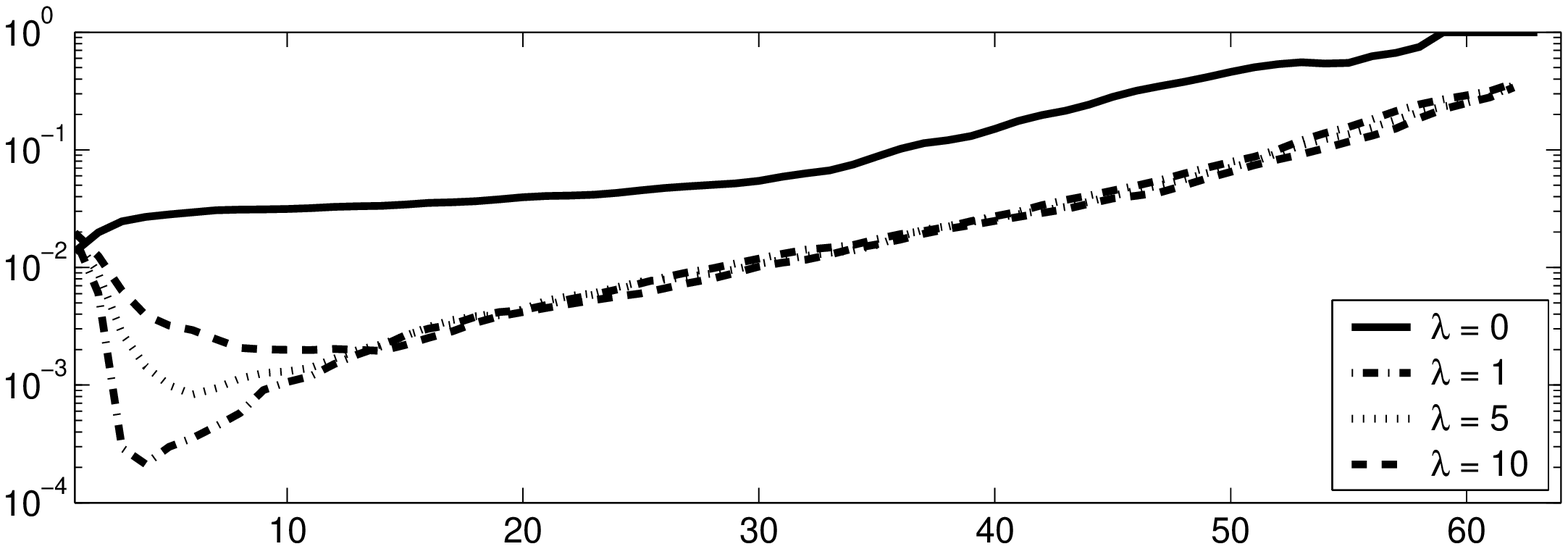}
		\caption{The mortality curves for different learning coefficients $\lambda$.}
	\label{fig:addRisk}
		\includegraphics[width=0.8\textwidth]{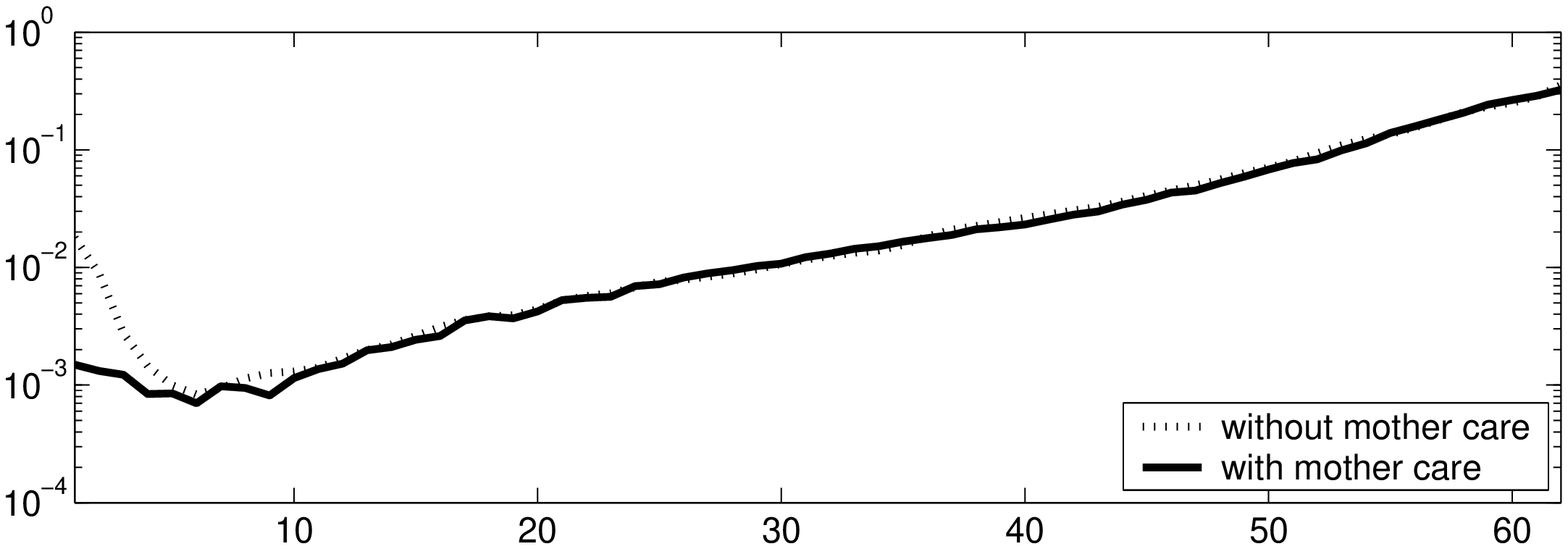}
		\caption{The mortality curve for learning coefficient $\lambda=5$ with and without mother care.}
	\label{fig:addRisk2}
		\includegraphics[width=0.8\textwidth]{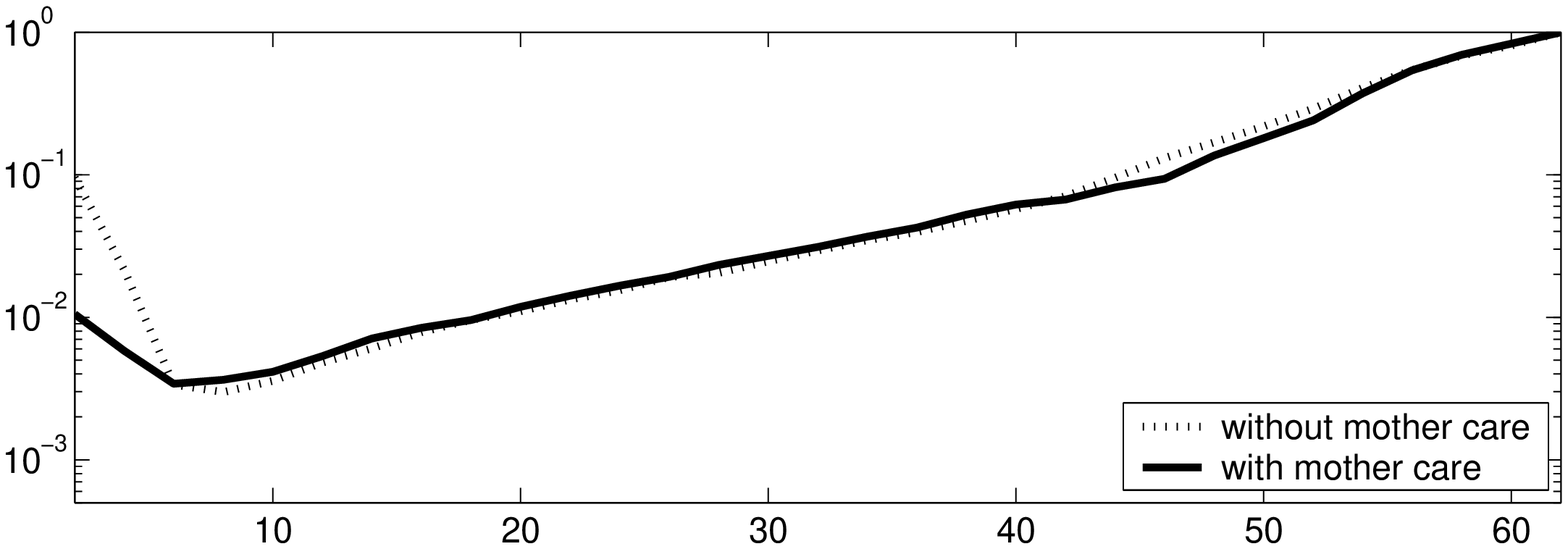}
		\caption{The mortality curve for populations with and without mother care. For early periods of live the mortality of newborns is lower in case of mother care, but then it is a bit higher.}
	\label{fig:addRisk3}
\end{figure}
The next question is how the intensive protection of newborns against environment fluctuations could influence their mortality during the later periods of life. It is rather well known effect that children who are very strongly protected against any infections during the first periods of life and live in almost sterile conditions are more vulnerable for infections later. Plot shown in Fig. 6 indicate, that one could really expect slightly higher mortality of young individuals if they are isolated from environment influence during the very early periods of life.

\section{Additional risk factors}

Suppose that after age $H$ in every year an individual may die with some very small probability due to the random death (e.g. in a car accident or any other death of young people released from the parental care). In Fig. 7 we presented results for populations, in which individuals older than $H=12$ dies in each year with probability $p=0.002$ even if their state is lover than the $F$. The mortality curve resembles the curves observed in many real human populations. This random death introduced for individuals older than 12 could be also natural increase in mortality connected with reaching the puberty age.  
In Fig. 8 we presented the age structure of different human populations and for comparison, the mortality curves generated by simulations with properly rescaled the age axis. 

\begin{figure}
	\centering
		\includegraphics[width=0.8\textwidth]{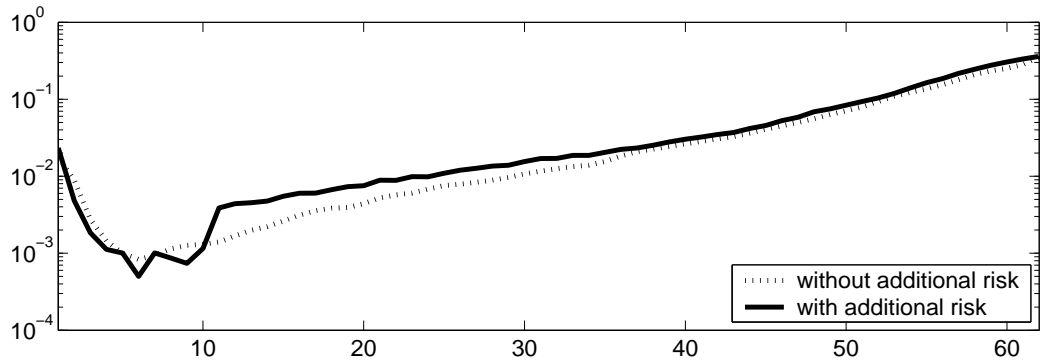}
		\caption{The mortality curve for the population with additional risk of death after $H=12$ years.}
	\label{fig:addRisk}
\end{figure}

\begin{figure}
	\centering
		\includegraphics[width=0.8\textwidth]{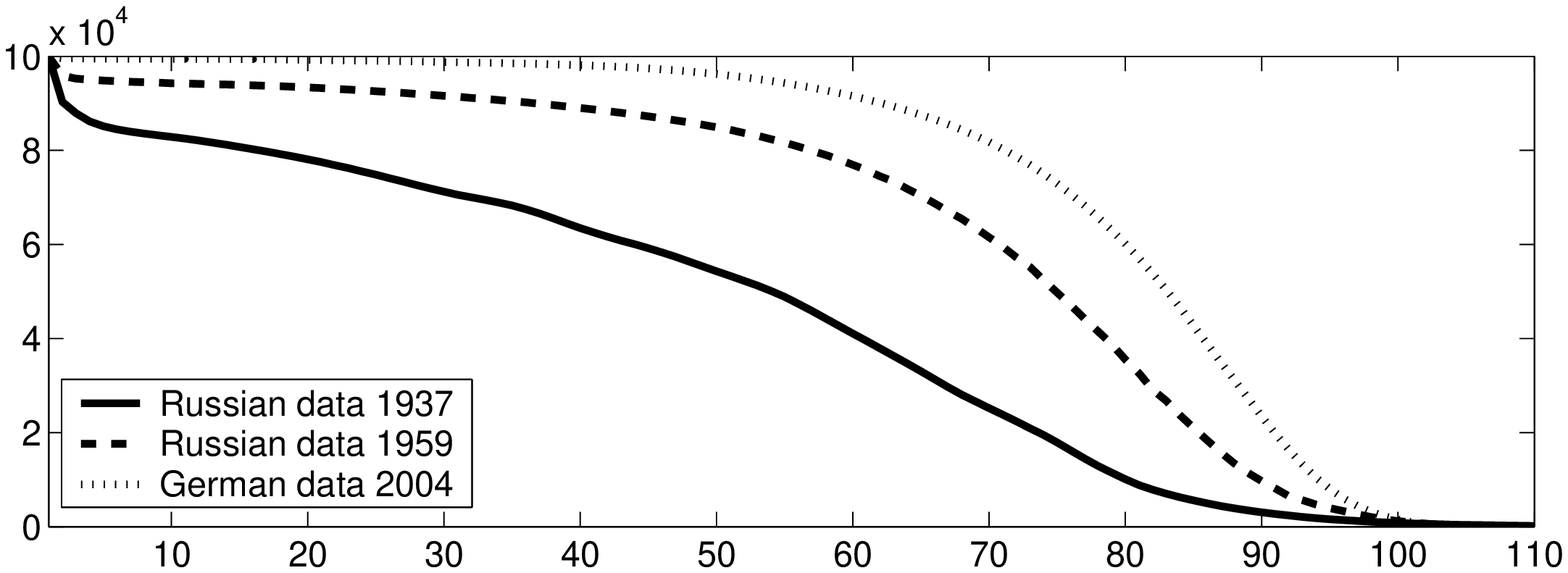}
		\includegraphics[width=0.8\textwidth]{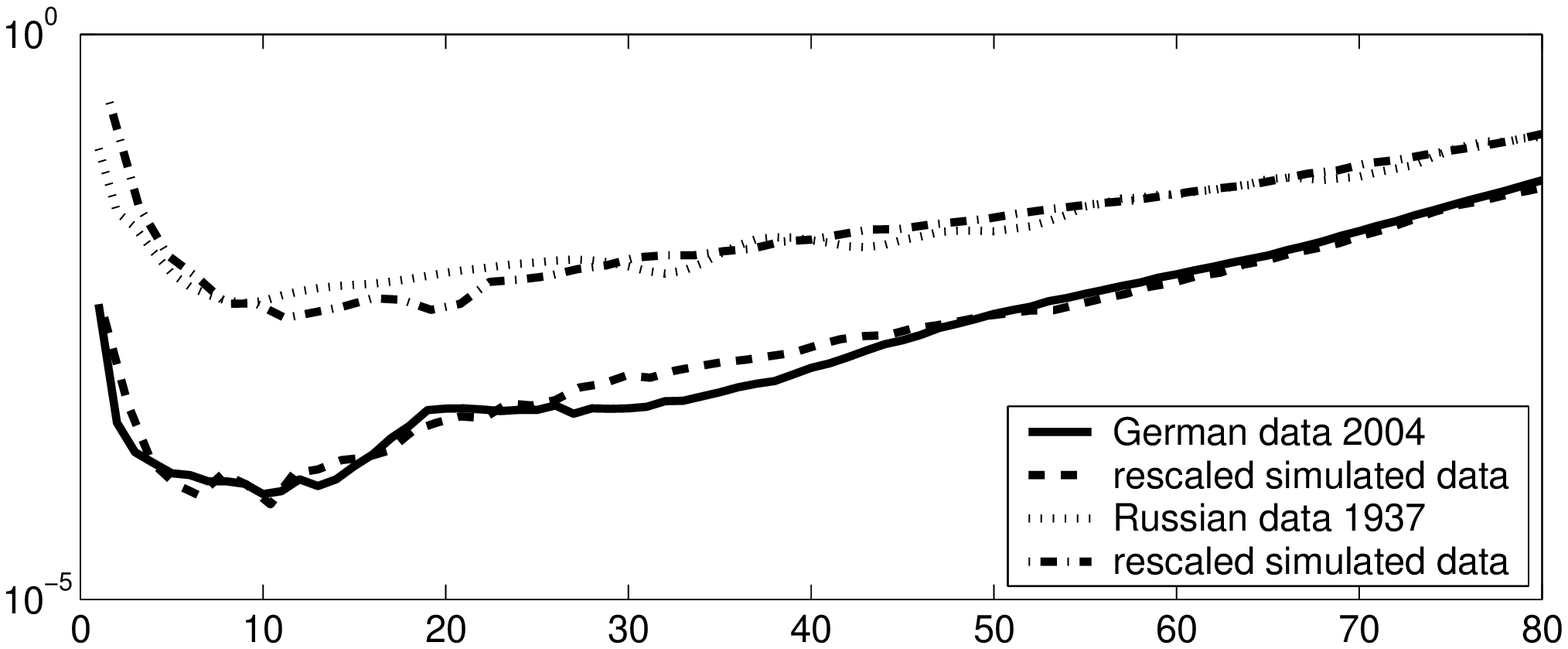}
		\caption{The age structures (upper plots) and mortality curves of different human populations and populations generated in simulations with the extended noisy Penna model (lower plots).}
	\label{fig:addRisk}
\end{figure}

\section{Conclusions}
The results of standard Penna model simulations reproduce the age distribution of the human population in its part after the minimum reproduction age. The noisy model enables simulation and analysis of parameters which influences the mortality of the youngest individuals. It will be possible to use this model for simulation the effect of pandemics and long distance immunological memory of populations.

\section{Acknowledgments}
Authors thank D. Stauffer for discussions. This work was supported by the grant 1 P03A 017 29, grant \# 105/E-344/SPB and Polish Foundation for Science. It was done in the frame of European programs: COST Action P10 and NEST - GIACS.

\end{document}